# Establishing Virtual R&D Teams: Obliged Policy

Nader Ale Ebrahim[1], Shamsuddin Ahmed and Zahari Taha

Department of Engineering Design and Manufacture,

Faculty of Engineering, University of Malaya, Kuala Lumpur, Malaysia

[1]Phone: +60-17-3942458, Fax: +60-7967-5330, Email: aleebrahim@perdana.um.edu.my

**Abstract:**

In a global and technology oriented world the requirements that products and services have to fulfill are increasing and are getting more complicated. Research and development (R&D) is becoming increasingly important in creating the knowledge that makes research and business more competitive. Companies are obliged to produce more rapidly, more effectively and more efficiently. In order to meet these requirements and to secure the viability of business processes, services and products R&D teams need to access and retrieve information from as many sources as possible. From the other perspective virtual teams are important mechanisms for organizations seeking to leverage scarce resources across geographic and other boundaries moreover; virtual collaboration has become vital for most organizations. This is particularly true in the context of designing new product and service innovation. Such collaboration often involves a network of partners located around the world. However at the R&D project level, dealing with such distributed teams challenges both managers and specialists. In new product development, it is necessary to put together the growing different capabilities and services with the goal, through cooperation between suppliers and customers, service providers and scientific institutions to achieve innovations of high quality.

In this paper based on comprehensive literature review of recent articles, at the first step provides an primary definition and characterization of virtual R&D team; next, the potential value created by virtual R&D teams for new product development is explored and lastly along with a guide line for future study, it is argued that the establishing of virtual R&D teams should be given consideration in the management of R&D projects.

**Introduction:**

The pressure of globalization competition force producers to continuously innovate and upgrade the quality of existing products (Acs and Preston, 1997) and organizations are currently facing important and unprecedented challenges in an ever dynamic, constantly changing and complex environment (Rezgui, 2007). In different point of view innovation is becoming the most important key issue for company's success in the 21st century (Sorli et al., 2006). The internationalization of R&D network is recent phenomenon (Kafouros et al., 2008). Considering the fact that in this knowledge-based environment, the driving forces for this phenomena are digitization, the internet, and high-speed data networks that are keys to addressing many of the operational issues from design to logistics and distribution (Noori and Lee, 2006). From the other direction to surviving in the highly competitive industry, requires strategies to collaborate with or compete with suitable firms within a network in the NPD process (Chen et al., 2008b).

The mega trends like globalization and high demand fluctuation force companies and supply chains to innovate new business models to gain and maintain competitive position. Networking, outsourcing, and information and communication technology are considered as general tools and means to respond to these challenges (Salmela and Lukka, 2004). As a consequence multinational enterprises have increased their research and development (R&D) investment in foreign countries (Reger, 2004). While the outsourcing activities of the MNCs was highly concentrated in a handful of economies by the beginning of the global R&D wave, the offshore outsourced R&D activities have now been more geographically dispersed and this indeed reveals the increasing value of networking and networks. These multiple sites encourage the development of more ideas, due to the varied international backgrounds in global networks (Richtne´r and Rognes, 2008).

In this paper the following aspects - comprehensive definition of virtual R&D teams, distributed work , achieve to pool of talent, importance of networking, new product development and virtuality, and advantages of applying virtual teams, are discussed in technical terms. Finally guide line for future studies; and establishing of virtual R&D teams is argued.

**Comprehensive definition of Virtual R&D teams:**

In this era popularity for virtual team structures in organizations is growing (Walvoord et al., 2008). Martins et al. (2004) in a major review of the literature on virtual teams, conclude that 'with rare exceptions all organizational teams are virtual to some extent.' Organizations have moved away from working with people who are in our visual proximity to working with people around the globe (Johnson et al., 2001). Although virtual teamwork is a current topic in the literature on global organizations but it has been problematic to define what is 'virtual' means across multiple institutional contexts (Chudoba et al., 2005). The concept of a "team" has been described as a small number of people with complementary skills who are equally committed to a common purpose, goals, and working approach for which they hold themselves mutually accountable (Zenun et al., 2007). It's a widely accepted fact that innovation is better achieved by working in team (Sorli et al., 2006). A majority of successful innovations is developed through the collective efforts of individuals in new product development teams (Akgun et al., 2006). All teams and virtual teams in particular, must develop mechanisms for sharing knowledge, experiences, and insights critical for accomplishing their missions (Rosen et al., 2007).

It is a worth mentioning that virtual teams are often formed to overcome geographical or temporal separations (Cascio and Shurygailo, 2003). Virtual teams work across boundaries of time and space by utilizing modern computer-driven technologies. The term "virtual team" is used to cover a wide range of activities and forms of technology-supported working (Anderson et al., 2007). Virtual teams are comprised of members who are located in more than one physical location. This team trait has fostered extensive use of a variety of forms of computer-mediated communication that enable geographically dispersed members to coordinate their individual efforts and inputs (Peters and Manz, 2007). (Gassmann and Von Zedtwitz, 2003b) defined "virtual team as a group of people and sub-teams who interact through interdependent tasks guided by common purpose and work across links strengthened by information, communication, and transport technologies." Another definition suggests that virtual teams are distributed work teams whose members are geographically dispersed and coordinate their work predominantly with electronic information and communication technologies (e-mail, video-conferencing, telephone, etc.) (Hertel et al., 2005). Different authors have identified diverse areas. From the perspective of Leenders et al. (2003) virtual teams are groups of individuals collaborating in the execution of a specific project while geographically and often temporally distributed, possibly anywhere within (and beyond) their parent organizations. Lurey and Raisinghani (2001) defined virtual teams - groups of people who work together although they are often dispersed across space, time, and/or organizational boundaries. Amongst the different definitions of a virtual team the following concept from which the term employed in this paper is one of the most widely accepted definition: (Powell et al., 2004), ''virtual teams are as groups of geographically, organizationally and/or time dispersed workers brought together by information technologies to accomplish one or more organization tasks ''.

**Distributed work:**

Companies are going global and this is especially true for companies participating in the global supply chain (Chen et al., 2007). Responding to the increasing de-centralization and globalization of work processes, many organizations have responded to their dynamic environments by introducing virtual teams. Virtual teams are growing in popularity (Cascio, 2000). Additionally, the rapid development of new communication technologies such as the Internet has accelerated this trend so that today, most of the larger organization employs virtual teams to some degree (Hertel et al., 2005). Taking into account that under the increasingly competitive global market, a firm simply cannot survive without new products developed under network cooperation, especially for high-tech industries (Chen et al., 2008a). Firms rely heavily on new product development to successfully compete in increasingly competitive global markets (Batallas and Yassine, 2004).

With the rapid development of electronic information and communication media in the last decades, distributed work has become much easier, faster and more efficient (Hertel et al., 2005). Now global communication is so much easier, faster and cheaper, therefore managing and integrating geographically dispersed R&D has considerably increased (Hegde and Hicks, 2008).
 Information technology is providing the infrastructure necessary to support the development of new organization forms. Virtual teams represent one such organizational form, one that could revolutionize the workplace and provide organizations with unprecedented level of flexibility and responsiveness (Powell et al., 2004). Moreover information and communication technology (ICT) has brought about significant changes in organizations and produced important benefits, including in the areas of marketing and innovation and many works highlight the importance of ICT as a key element in integrating marketing into the NPD process (Vilaseca-Requena et al., 2007). The employed Web Services technology, although very popular nowadays but it is still not mature enough, so dealing with it can bring new findings (Witczynski, 2006). Considering that R&D teams need to access and retrieve information from as many sources as possible (Kafouros et al., 2008), virtual teams are important mechanisms for organizations seeking to leverage scarce resources across geographic and other boundaries (Munkvold and Zigurs, 2007).The global competition and accelerated improvements in basic technologies demand organizations to develop the ability to manage efficient NPD projects that yield innovative products (Naveh, 2005). It's a widely held belief that the modern work-world is dominated by computer-mediated communication, and this communication is the bread and butter of virtual teams (Walvoord et al., 2008). In other words advancement in technologies and management skills has blurred firm boundaries (Acs and Preston, 1997).

**Achieve to pool of talent:**

Internationalization of markets, specialization of skills and knowledge, and the requirement to involve an increasingly large pool of knowledge simultaneously in the NPD process have all pushed firms to rely more and more on dispersed teams in their NPD endeavor (Leenders et al., 2003). In the past an original equipment manufacturer company (OEM) would have a dominant role with its suppliers, determining the specifications for requirements and waiting for suppliers to compete for orders to supply the required components. Now complex products are designed much more collaboratively with the suppliers being involved in the design process. The production of a new car for example involves different companies in the supply chain acting more as partners in a joint manufacturing exercise (Anderson et al., 2007). However by comparison in today's competitive global economy, organizations capable of rapidly creating virtual teams of talented people can respond quickly to changing business environments. capabilities of this type offer organizations a form of competitive advantage (Bergiel et al., 2008).

**Importance of Networking:**

As another important aspects shedding light on the importance of networking it can be mentioned that since the maximum profit of the network can be obtained by sharing the risk and the benefit with participants, it is important for corporations to collaborate in networks in order to develop capacity, capability and competence to perform new product development and become suppliers of complete systems (Chen et al., 2008b). Many R&D projects addressed already the issue of computer supported source networks (Witczynski, 2006). As another milestone phenomenon, over the last decades, R&D teams have become increasingly virtual (Kratzer et al., 2006). Virtual teams have become critical for companies to survive (Lurey and Raisinghani, 2001) The main advantage of implementing a geographically dispersed R&D network structure is the ability to tap selectively into center of excellence (Criscuolo, 2005).To shrink the cost and protracted length of total system and product development life cycles, many organizations have moved away from serial to concurrent collaboration through the use of cross-functional, integrated project/product teams (Bochenek and Ragusa, 2004). In addition to this such learning networks can generate localized social capital and endogenous growth dynamics (Cheng et al., 2001, Conceicao and Heitor, 2007).

It is clear that a team's behaviors in adopting the technology and adapting their processes will be influenced by the feature-set of the computer-mediated collaborations (CMCs) environment that is used (Rice et al., 2007).The growing complexity and competition in the business world are major drivers for increasing the popularity and formation of virtual teams (Chen et al., 2007).From a different perspective virtual teams are formed to facilitate transnational innovation processes (Gassmann and Von Zedtwitz, 2003b) and it should be noted that innovation has a positive impact on corporate performance (Kafouros et al., 2008). Also a virtual network structure is used to improve communication and coordination, and encourage the mutual sharing of inter-organizational resources and competencies (Chen et al., 2008b). Hance virtuality seems well suited for cultivating and managing creativity in NPD teams (Leenders et al., 2003).

**New Product development and virtuality:**

The life cycle of a product/good becomes shorter every year. Today, leading-edge firms can exploit global asset configurations to customize the existing products and services. They also have the ability to combine their resources with an expanding knowledge-base to create a continuous stream of new products and services (Miles et al., 2000). With the needs to respond quickly to dynamic customer demands, increasing complexity of product design and rapidly changing technologies, the selection of the right set of NPD is critical to a company's long-term success (Chen et al., 2008b). Also, combination of factors such as ever changing market needs and expectations, rough competition and emerging technologies among others, challenges being faced by industrial companies to continuously increase the rate of new products to the market to fulfill all these requirements (Sorli et al., 2006). The ultimate objective of all NPD teams is to acquire superior marketplace through new products (Akgun et al., 2006). In light of the above, product innovation is the central force in securing a firm's competitive advantage in international markets (Jeong, 2003). Therefore, NPD is vital and needs to be developed both innovatively and steadily (Chen et al., 2008b).

A multidisciplinary approach is needed to be successful in launching new products and managing daily operations (Flores, 2006). In NPD context, teams developing new products in turbulent environments encounter quick depreciation of technology and market knowledge due to rapidly changing customer needs, wants, and desires, and technological know-how (Akgun et al., 2007). ICT helps establish and maintain communication and cooperative relationships both inside and outside the organization, and makes NPD processes quicker, simpler and less risky (Vilaseca-Requena et al., 2007). ICT enhance the NPD process by shortening distances and saving on costs and time (Vilaseca-Requena et al., 2007). Various studies also offered a large number of examples from the industry showing how firms have been using the Internet in their NPD activities (Ozer, 2004, Ozer, 2000). Moreover, several recent studies specifically dealt with the development of new technologies and their impact on new product development among globally dispersed teams (McDonough et al., 2001, Jeong, 2003). Competitive strategies are forcing companies to deploy their NPD resources globally and, thus making collocated NPD teams prohibitively expensive and logistically difficult to manage (Susman et al., 2003)**.**

**Advantages of applying virtual teams:**

Virtual teams often face tight schedules and a need to start quickly and perform instantly (Munkvold and Zigurs, 2007). Virtual team may allow people to collaborate more productivity at a distance, but the tripe to coffee corner or across the hallway to a trusted colleague is still the most reliable and effective way to review and revise a new idea (Gassmann and Von Zedtwitz, 2003a). Virtual teams reduce time-to-market (May and Carter, 2001). Lead time or time to market has been generally admitted to be one of the most important keys for success in manufacturing companies (Sorli et al., 2006). In a virtual team environment, collaborative and competitive conflicting behavior is positively linked with performance (Powell et al., 2004), depending on the degree of virtuality (Ortiz de Guinea et al., 2005) and team connectivity (Ortiz de Guinea et al., 2005). As drawbacks, virtual teams are particularly vulnerable to mistrust, communication break downs, conflicts, and power struggles (Rosen et al., 2007). Table 1 summarizes some of the main advantages associated with virtual teaming.

Table 1: Main advantages associated with virtual teaming

| Advantages | Reference |
| --- | --- |
| Reduce relocation time and costs, reduced travel costs, Greater productivity, shorter development times. | (McDonough et al., 2001, Rice et al., 2007, Bergiel et al., 2008, Cascio, 2000, Fuller et al., 2006) |
| Virtual teams reduce time-to-market | (May and Carter, 2001) |
| Ability to digitally or electronically unite experts in highly specialized fields working at great distances from each other | (Rosen et al., 2007) |
| Have more effective R&D continuation decisions, Most effective in making decisions. | (Cummings and Teng, 2003) (Hossain and Wigand, 2004) |
| Ability to tap selectively into center of excellence, using the best talent regardless of location, Allow organizations to access the most qualified individuals for a particular job regardless of their location. | (Criscuolo, 2005, Cascio, 2000, Samarah et al., 2007, Fuller et al., 2006) (Gassmann and Von Zedtwitz, 2003b) (Munkvold and Zigurs, 2007) (Hunsaker and Hunsaker, 2008) |
| Greater degree of freedom to individuals involved with the development project | (Ojasalo, 2008) |
| Producing better outcomes and attract better employees | (Martins et al., 2004, Rice et al., 2007) |
| Provide flexible working hours for the employees, Create and disperse improved business processes across organizations, Do a good job and finish their work on time, Change the direction of the company from a production-oriented company to a service/information-oriented company and allow organizations to hire and retain | (Johnson et al., 2001), (Paul et al., 2005 ), (Precup et al., 2006) |
| people, regardless of location. Resistance to change is reduced. Faster response times to tasks, Provide a vehicle for global collaboration and coordination of R&D-related activities. | |
| Useful for projects that require cross-functional or cross boundary skilled inputs | (Lee-Kelley and Sankey, 2008) |
| Teams can be organized whether or not members are in proximity to one another | (Kratzer et al., 2005, Cascio, 2000) |
| Provide organizations with unprecedented level of flexibility and responsiveness | (Powell et al., 2004, Hunsaker and Hunsaker, 2008) |
| Perform their work without concern of space or time constraints | (Lurey and Raisinghani, 2001) |
| Self-assessed performance. | Chudoba et al. (2005) |
| Optimize the contributions of individual members toward the completion of business tasks and organizational goal | (Samarah et al., 2007) |
| reduce the pollution | (Johnson et al., 2001) |
| The ratio of virtual R&D member publications exceeded from co-located publications | (Ahuja et al., 2003) |
| Extent of informal exchange of information is minimal | (Pawar and Sharifi, 1997) |
| Can manage the development and commercialization tasks quite well | (Chesbrough and Teece, 2002) |
| Facilitate transnational innovation processes | (Gassmann and Von Zedtwitz, 2003b) |
| Respond quickly to changing business environments | (Bergiel et al., 2008) |
| Improve communication and coordination, and encourage the mutual sharing of inter-organizational resources and competencies | (Chen et al., 2008b) |
| Team communications and work reports are available online to facilitate swift responses to the demands of a global market. Employees can be assigned to multiple, concurrent teams; dynamic team membership allows people to move from one project to another. Employees can more easily accommodate both personal and professional lives. | (Cascio, 2000) |
| Cultivating and managing creativity | (Leenders et al., 2003) |
| Sharing knowledge, experiences | (Rosen et al., 2007, Zakaria et al., 2004) |
| Improve the detail and precision of design activities | (Vaccaro et al., 2008) |
| Enable organizations to respond faster to increased competition | (Hunsaker and Hunsaker, 2008, Pauleen, 2003) |
| Better team outcomes (quality, productivity, and satisfaction) | (Gaudes et al., 2007 , Ortiz de Guinea et al., 2005) |
| | |
| Higher team effectiveness and efficiency | (May and Carter, 2001, Shachaf and Hara, 2005) |

## Conclusion:

This paper provides a comprehensive literature review of virtual R&D team, based upon recent articles. Despite the massive benefaction of virtual R&D team which is described in table 2, the application of virtual team to upgrade and enhance business operation by most enterprises, is still at its infancy. While reviewing the previous study, it is believed that the advantages of working on the basis of virtual teams far outweigh the disadvantages. Virtual teams bring about knowledge spillovers within enterprises bridging time and place, reduce time-to-market, reduced travel costs, ability to tap selectively into center of excellence, using the best talent regardless of location, greater degree of freedom to individuals, shorter development times, provide flexible hours for the employees the working hours, creates and disperses improved business processes across organizations, provide organizations with unprecedented level of flexibility and responsiveness, reduce resistance to change, reduce the pollution, Optimize the contributions of individual members toward the completion of business tasks and organizational goal, facilitate transnational innovation processes, respond quickly to changing business environments, employees can be assigned to multiple, concurrent teams and finally higher team effectiveness and efficiency. Therefore the decision on setting up virtual teams is not a choice but a requirement.

Future research need to design infrastructures to support virtual R&D team working. New ways of communicating and interacting among team members in virtual environments will necessitate to be developed and implemented .Innovative business trend, involved with burden by workers for more flexibility and empowerment, suggest establishing virtual R&D Teams for new product development which is obligation policy for companies managers.

**References:**

ACS, Z. J. & PRESTON, L. (1997) Small and Medium-Sized Enterprises, Technology, and Globalization: Introduction to a Special Issue on Small and Medium-Sized Enterprises in the Global Economy. *Small Business Economics,* 9**,** 1-6.
AHUJA, M. K., GALLETTA, D. F. & CARLEY, K. M. (2003) Individual Centrality and Performance in Virtual R&D Groups: An Empirical Study *Management Science,* 49**,** 21-38.
AKGUN, A. E., BYRNE, J. C., LYNN, G. S. & KESKIN, H. (2007) New product development in turbulent environments: Impact of improvisation and unlearning on new product performance. *Journal of Engineering and Technology Management,* 24**,** 203–230.
AKGUN, A. E., LYNN, G. S. & YILMAZ, C. (2006) Learning process in new product development teams and effects on product success: A socio-cognitive perspective. *Industrial Marketing Management,* 35**,** 210 – 224.
ANDERSON, A. H., MCEWAN, R., BAL, J. & CARLETTA, J. (2007) Virtual team meetings: An analysis of communication and context. *Computers in Human Behavior,* 23**,** 2558–2580.
BATALLAS, D. A. & YASSINE, A. A. (2004) Information Leaders in Product Development Organizational Networks: Social Network Analysis of the Design Structure Matrix. *Presented at "Understanding Complex Systems" Symposium.* Urbana-Champaign, University of Illinois.
BERGIEL, J. B., BERGIEL, E. B. & BALSMEIER, P. W. (2008) Nature of virtual teams: a summary of their advantages and disadvantages. *Management Research News,* 31**,** 99-110.
BOCHENEK, G. & RAGUSA, J. (2004) Improving Integrated Project Team Interaction Through Virtual (3D) Collaboration. *Engineering Management Journal,* 16**,** 3.
CASCIO, W. F. (2000) Managing a virtual workplace. *The Academy of Management Executive,* 14**,** 81-90.
CASCIO, W. F. & SHURYGAILO, S. (2003) E-Leadership and Virtual Teams. *Organizational Dynamics,* 31**,** 362-376.
CHEN, H. H., KANG, Y. K., XING, X., LEE, A. H. I. & TONG, Y. (2008b) Developing new products with knowledge management methods and process development management in a network. *Computers in Industry,* 59**,** 242–253.
CHEN, H. H., LEE, A. H. I., WANG, H. Z. & TONG, Y. (2008a) Operating NPD innovatively with different technologies under a variant social environment. *Technological Forecasting & Social Change***,** 385–404.
CHEN, M., LIOU, Y., WANG, C. W., FAN, Y. W. & CHI, Y. P. J. (2007) Team Spirit: Design, implementation, and evaluation of a Web-based group decision support system. *Decision Support Systems,* 43**,** 1186–1202.
CHENG, E. W. L., LI, H., LOVE, P. E. D. & IRANI, Z. (2001) An e-business model to support supply chain activities in construction. *Logistics Information Management,* 14**,** 68-77.
CHESBROUGH, H. W. & TEECE, D. J. (2002) Organizing for Innovation: When Is Virtual Virtuous? *Harvard Business Review Article,* August 127-135.
CHUDOBA, K. M., WYNN, E., LU, M., WATSON-MANHEIM & BETH, M. (2005) How virtual are we? Measuring virtuality and understanding its impact in a global organization. *Information Systems Journal,* 15**,** 279-306.

CONCEICAO, P. & HEITOR, M. V. (2007) Diversity and integration of science and technology policies. *Technological Forecasting & Social Change,* 74**,** 1–17.
CRISCUOLO, P. (2005) On the road again: Researcher mobility inside the R&D network. *Research Policy,* 34**,** 1350–1365
CUMMINGS, J. L. & TENG, B. S. (2003) Transferring R&D knowledge: the key factors affecting knowledge transfer success. *Journal of Engineering Technology Management***,** 39–68.
FLORES, M. (2006) IFIP International Federation for Information Processing. *Network-Centric Collaboration and Supporting Fireworks.* Boston, Springer.
FULLER, M. A., HARDIN, A. M. & DAVISON, R. M. (2006) Efficacy in Technology-Mediated Distributed Team *Journal of Management Information Systems,* 23**,** 209-235.
GASSMANN, O. & VON ZEDTWITZ, M. (2003a) *Innovation Processes in Transnational Corporations*, Elsevier Science Ltd.
GASSMANN, O. & VON ZEDTWITZ, M. (2003b) Trends and determinants of managing virtual R&D teams. *R&D Management* 33**,** 243-262.
GAUDES, A., HAMILTON-BOGART, B., MARSH, S. & ROBINSON, H. (2007 ) A Framework for Constructing Effective Virtual Teams *The Journal of E-working* 1**,** 83-97
HEGDE, D. & HICKS, D. (2008) The maturation of global corporate R&D: Evidence from the activity of U.S. foreign subsidiaries. *Research Policy,* 37**,** 90–406.
HERTEL, G. T., GEISTER, S. & KONRADT, U. (2005) Managing virtual teams: A review of current empirical research. *Human Resource Management Review,* 15**,** 69–95.
HOSSAIN, L. & WIGAND, R. T. (2004) ICT Enabled Virtual Collaboration through Trust. *Journal of Computer-Mediated Communication,* 10.
HUNSAKER, P. L. & HUNSAKER, J. S. (2008) Virtual teams: a leader's guide. *Team Performance Management,* 14**,** 86-101.
JEONG, I. (2003) A cross-national study of the relationship between international diversification and new product performance. *International Marketing Review,* 20**,** 353-376.
JOHNSON, P., HEIMANN, V. & O'NEILL, K. (2001) The "wonderland" of virtual teams. *Journal of Workplace Learning,* 13**,** 24 - 30.
KAFOUROS, M. I., BUCKLEY, P. J., SHARP, J. A. & WANG, C. (2008) The role of internationalization in explaining innovation performance. *Technovation,* 28**,** 63–74.
KRATZER, J., LEENDERS, R. & ENGELEN, J. V. (2005) Keeping Virtual R&D Teams Creative. *Industrial Research Institute, Inc.,* March-April**,** 13-16.
KRATZER, J., LEENDERS, R. T. A. J. & ENGELEN, J. M. L. V. (2006) Managing creative team performance in virtual environments: an empirical study in 44 R&D teams. *Technovation,* 26**,** 42–49.
LEE-KELLEY, L. & SANKEY, T. (2008) Global virtual teams for value creation and project success: A case study. *International Journal of Project Management* 26**,** 51–62.
LEENDERS, R. T. A. J., ENGELEN, J. M. L. V. & KRATZER, J. (2003) Virtuality, communication, and new product team creativity: a social network perspective. *Journal of Engineering and Technology Management,* 20**,** 69–92.
LUREY, J. S. & RAISINGHANI, M. S. (2001) An empirical study of best practices in virtual teams *Information & Management,* 38**,** 523-544.
MARTINS, L. L., GILSON, L. L. & MAYNARD, M. T. (2004) Virtual teams: What do we know and where do we go from here? *Journal of Management,* 30**,** 805–835.
MAY, A. & CARTER, C. (2001) A case study of virtual team working in the European automotive industry. *International Journal of Industrial Ergonomics,* 27**,** 171-186.
MCDONOUGH, E. F., KAHN, K. B. & BARCZAK, G. (2001) An investigation of the use of global, virtual, and collocated new product development teams. *The Journal of Product Innovation Management,* 18**,** 110–120.
MILES, R. E., SNOW, C. C. & MILES, G. (2000) TheFuture.org *Long Range Planning,* 33**,** 300-321.

MUNKVOLD, B. E. & ZIGURS, I. (2007) Process and technology challenges in swift-starting virtual teams. *Information & Management,* 44**,** 287–299.
NAVEH, E. (2005) The effect of integrated product development on efficiency and innovation. *International Journal of Production Research,* 43**,** 2789–2808.
NOORI, H. & LEE, W. B. (2006) Dispersed network manufacturing: adapting SMEs to compete on the global scale. *Journal of Manufacturing Technology Management,* 17.
OJASALO, J. (2008) Management of innovation networks: a case study of different approaches. *European Journal of Innovation Management,* 11**,** 51-86.
ORTIZ DE GUINEA, A., WEBSTER, J. & STAPLES, S. ( 2005) A Meta-Analysis of the Virtual Teams Literature. *Symposium on High Performance Professional Teams Industrial Relations Centre.* School of Policy Studies, Queen's University, Kingston, Canada.
OZER, M. (2000) Information Technology and New Product Development Opportunities and Pitfalls. *Industrial Marketing Management* 29**,** 387-396.
OZER, M. (2004) The role of the Internet in new product performance: A conceptual investigation. *Industrial Marketing Management* 33**,** 355– 369.
PAUL, S., SEETHARAMAN, P., SAMARAH, I. & PETER MYKYTYN, J. ( 2005 ) Understanding Conflict in Virtual Teams: An Experimental Investigation using Content Analysis. *38th Hawaii International Conference on System Sciences.* Hawaii.
PAULEEN, D. J. (2003) An Inductively Derived Model of Leader-Initiated Relationship Building with Virtual Team Members. *Journal of Management Information Systems,* 20**,** 227-256.
PAWAR, K. S. & SHARIFI, S. (1997) Physical or virtual team collocation: Does it matter? *International Journal of Production Economics* 52**,** 283-290.
PETERS, L. M. & MANZ, C. C. (2007) Identifying antecedents of virtual team collaboration. *Team Performance Management,,* 13**,** 117-129.
POWELL, A., PICCOLI, G. & IVES, B. (2004) Virtual teams: a review of current literature and directions for future research. *The Data base for Advances in Information Systems,* 35**,** 6–36.
PRECUP, L., O'SULLIVAN, D., CORMICAN, K. & DOOLEY, L. (2006) Virtual team environment for collaborative research projects. *International Journal of Innovation and Learning,* 3**,** 77 - 94
REGER, G. (2004) Coordinating globally dispersed research centers of excellence—the case of Philips Electronics. *Journal of International Management,* 10**,** 51– 76.
REZGUI, Y. (2007) Exploring virtual team-working effectiveness in the construction sector. *Interacting with Computers,* 19**,** 96–112.
RICE, D. J., DAVIDSON1, B. D., DANNENHOFFER, J. F. & GAY, G. K. (2007) Improving the Effectiveness of Virtual Teams by Adapting Team Processes. *Computer Supported Cooperative Work,* 16**,** 567–594.
RICHTNÉR, A. & ROGNES, J. (2008) Organizing R&D in a global environment-Increasing dispersed co-operation versus continuous centralization. *European Journal of Innovation Management,* 11.
ROSEN, B., FURST, S. & BLACKBURN, R. (2007) Overcoming Barriers to Knowledge Sharing in Virtual Teams. *Organizational Dynamics,* 36**,** 259–273.
SALMELA, E. & LUKKA, A. (2004) Value added logistics in supply and demand chains SMILE. Part 1 : Ebusiness between global company and its local SME supplier network, Research Report 153, ISBN 951-764-925-8.
SAMARAH, I., PAUL, S. & TADISINA, S. (2007) Collaboration Technology Support for Knowledge Conversion in Virtual Teams: A Theoretical Perspective. *40th Hawaii International Conference on System Sciences (HICSS).* Hawai.
SHACHAF, P. & HARA, N. (2005) Team Effectiveness in Virtual Environments: An Ecological Approach. IN FERRIS, P. A. G., S., (Ed.) *Teaching and Learning with Virtual Teams.* Idea Group Publishing.
SORLI, M., STOKIC, D., GOROSTIZA, A. & CAMPOS, A. (2006) Managing product/process knowledge in the concurrent/simultaneous enterprise environment. *Robotics and Computer-Integrated Manufacturing,* 22**,** 399–408.

SUSMAN, G. I., GRAY, B. L., PERRY, J. & BLAIR, C. E. (2003) Recognition and reconciliation of differences in interpretation of misalignments when collaborative technologies are introduced into new product development teams. *Journal of Engineering and Technology Management,* 20**,** 141–159.

VACCARO, A., VELOSO, F. & BRUSONI, S. (2008) The Impact of Virtual Technologies on Organizational Knowledge Creation: An Empirical Study. *Hawaii International Conference on System Sciences.* Proceedings of the 41st Annual Publication

VILASECA-REQUENA, J., TORRENT-SELLENS, J. & JIMÉNEZ-ZARCO, A. I. (2007) ICT use in marketing as innovation success factor-Enhancing cooperation in new product development processes. *European Journal of Innovation Management,* 10**,** 268-288.

WALVOORD, A. A. G., REDDEN, E. R., ELLIOTT, L. R. & COOVERT, M. D. (2008) Empowering followers in virtual teams: Guiding principles from theory and practice", Computers in Human Behavior (article in press).

WITCZYNSKI, M. (2006) Network-Centric Collaboration and Supporting Fireworks. IN CAMARINHA-MATOS, L., AFSARMANESH, H. & OLLUS, M. (Eds.) *IFIP International Federation for Information Processing.* Boston, Springer.

ZAKARIA, N., AMELINCKX, A. & WILEMON, D. (2004) Working Together Apart? Building a Knowledge-Sharing Culture for Global Virtual Teams. *Creativity and Innovation Management,* 13**,** 15-29.

ZENUN, M. M. N., LOUREIRO, G. & ARAUJO, C. S. (2007) The Effects of Teams' Co-location on Project Performance. IN LOUREIRO, G. & CURRAN, R. (Eds.) *Complex Systems Concurrent Engineering-Collaboration, Technology Innovation and Sustainability.* London, Springer.